\documentclass[aps,prl,twocolumn,amsmath,amssymb]{revtex4-1}
\usepackage[colorlinks,pdfusetitle,urlcolor=blue,citecolor=blue,linkcolor=blue,bookmarksnumbered,plainpages=false]{hyperref}
\usepackage{graphicx}
\newcommand{\B}{\mathbf}

\usepackage{verbatim}
\newcommand{\tens}{\B}
\renewcommand{\d}{\!\mathrm{d}}
\renewcommand{\i}{\mathrm{i}}

\usepackage{bm}
\usepackage{braket}
\usepackage[percent]{overpic}

\begin{document}

\title{Superradiance and symmetry in resonant energy transfer}
\author{Severin Bang}
\author{Stefan Yoshi Buhmann}
\author{Robert Bennett}
\affiliation{Physikalisches Institut, Albert-Ludwigs-Universit\"at Freiburg, Hermann-Herder-Str. 3, D-79104 Freiburg i. Br., Germany}
\date{\today}

\begin{abstract} Closely-spaced quantum emitters coherently sharing excitation can release their energy faster than suggested by a simple sum over their individual emission rates --- a phenomenon known as superradiance. Here, we show that the assumption of closely-spaced emitters can be relaxed in the context of resonant energy transfer, instead finding that certain symmetrical arrangements of donors are just as effective. We derive exact expressions for the superradiant fidelity in such situations, finding some surprising results such as complete suppression of the rate for a single acceptor within a homogeneous spherically symmetric distribution of coherent donors. \end{abstract}

\maketitle

When quantum emitters share an excitation and interact coherently, the stored energy can be released more quickly than it would be if the emitters were taken as independent. This phenomenon is known as superradiance, and is one of the most prominent examples of collective effects in quantum theory. First predicted in 1954 \cite{Dicke1954}, its first experimental verification was carried out in sodium vapour \cite{Gross1976} and has been numerously observed since. Its most convenient theoretical description is based on the Dicke states, where the ensemble of two-level emitters is mapped to the algebra of angular momentum (see \cite{Gross1982} for a review). The criterion for superradiant behaviour in this context is that the atom--atom separation should be significantly smaller than the wavelength of the involved transitions.

The energy released from a decaying system can also be captured by another system, causing the latter to reach an excited state. This is known as (F\"{o}rster) resonant energy transfer (FRET/RET) \cite{Forster1946}, and is a pervasive mechanism by which energy can be transported from donor to acceptor (for recent reviews, see \cite{Govorov2016,Salam2018,Jones2019b}). Its characteristic inverse sixth power distance-dependence in the near field is a widely-used `spectroscopic ruler' in the analysis of macromolecular structure, while analogous processes such as interatomic Coulombic decay \cite{Cederbaum1997} are the subject of intense interest due to their relevance in radiation biology \cite{Gokhberg2014}. A description beyond the electrostatic dipole--dipole regime was developed on the basis of molecular quantum electrodynamics (QED), revealing that the resonance energy transfer rate beyond the nonretarded F\"orster regime falls off with an inverse quadratic distance law \cite{Andrews1989}. Many-particle effects in the form of passive mediator atoms have been shown to modify the transfer \cite{Craig1989,Salam2012b,Ford2014}. On a macroscopic level, multilayer structures can have a similar effect \cite{Biehs2016,Deshmukh2018}. 

Generalised F\"{o}rster theory is a many-particle version of FRET where the excitations mediating the process can be spread over a collection of donors or acceptors making up a molecular aggregate  \cite{Sumi1998,Mukai1999,Scholes2000,Scholes2001}. Superradiant behaviour is a signature of this arrangement, usually expressed as delocalisation of a quasiparticle excitation known as an exciton. Here we will show that in contrast to spontaneous decay, achieving a superradiant FRET rate does not require closely-spaced donors or acceptors. We will do this by building up a general expression for the superradiant fidelity, taking into account arbitrary arrangements of donors and acceptors as well as both near- and far-field behaviour. 

We work within the general formalism of macroscopic QED, making our work amenable to inclusion of dielectric bodies near the decaying cloud. Within this framework, the transfer rate $\Gamma$ from a single donor at $\B{r}_\text{D}$ to a single acceptor at $\B{r}_\text{A}$ is given by \cite{Dung2002};
\begin{equation}\label{GammaSingle}
\Gamma = \frac{2 \pi \mu_0^2 \omega^2}{\hbar} |\B{d}_\text{A} \cdot \tens{G}(\B{r}_\text{A},\B{r}_\text{D},\omega) \cdot \B{d}_\text{D}|^2
\end{equation}
where $\B{d}_\text{D}$ and $\B{d}_\text{A}$ are the transition dipole moments of donor and acceptor, respectively. The spectral overlap of donor and acceptor is taken here to be dominated by a single frequency, denoted as $\omega$. The matrix $\B{G}(\B{r},\B{r}',\omega)$ is the dyadic Green's tensor describing propagation of electromagnetic radiation of frequency $\omega = k/\sqrt{\varepsilon}$ from $\B{r}'$ to $\B{r}$ in a given environment, being defined as the solution to
\begin{equation}\label{GDefinition}
\nabla \times \nabla \times \tens{G}(\B{r},\B{r}',\omega) - {k^2}\tens{G}(\B{r},\B{r}',\omega) = \mathbb{I}\delta(\B{r}-\B{r}')\,. 
\end{equation}
supplemented by appropriate boundary conditions. The derivation of \eqref{GammaSingle} was based on Fermi's golden rule 
\begin{equation}\label{FGR}
\Gamma = \frac{2\pi}{\hbar} |M_{fi}|^2 \delta(E_i-E_f)
\end{equation}
with the transition matrix element
\begin{equation}
M_{fi} = \bra{f}_\text{A}\otimes \bra{f}_\text{D}H_\text{int}\ket{i}_\text{D}\otimes \ket{i}_\text{A}
\end{equation}
of the dipole coupling interaction Hamiltonian
\begin{equation}
H_\text{int} =-\hat{\B{d}}_\text{D} \cdot \hat{\B{E}}(\B{r}_\text{D}) -\hat{\B{d}}_\text{A} \cdot \hat{\B{E}}(\B{r}_\text{A}) 
\end{equation}
written in terms of the macroscopic QED electric field \cite{Gruner1996a, Buhmann2013} 
\begin{align}
\hat{\B{E}}(\B{r}) = \i \int_0^\infty \!\!\!\d \omega \frac{\omega^2}{c^2}&  \int \d^3 \B{r}' \sqrt{\frac{\hbar}{\pi \epsilon_0}\,\text{Im} \varepsilon(\B{r}',\omega)} \notag \\
&\times  \B{G}(\B{r}, \B{r}',\omega) \cdot \hat{\B{f}}(\B{r}',\omega) + \text{H.c.}
\end{align}
where $\hat{\B{f}}^\dagger$ and $\hat{\B{f}}$ are the creation and annihilation operators for elementary polariton-like excitations of the composite field-matter system, and $\varepsilon(\B{r},\omega)$ is the relative permittivity at position $\B{r}$ and frequency $\omega$. The important assumption here was that the donor and acceptor initial states $\ket{i}_\text{D}$ and $\ket{i}_\text{A}$ were simply the eigenstates $\ket{e}$ and $\ket{g}$ corresponding to the upper and lower levels in the involved transition, respectively. If the excitation is instead coherently spread over $N$ sites, we have;
 \begin{align}
\ket{i}_\text{D} =\frac{1}{\sqrt{N}} \left(\ket{e}_1+\ket{e}_2+\ldots+ \ket{e}_{\!N}\right)
\end{align}
one can follow the same steps as the previous calculation to find
\begin{align}\label{Gammaij}
\Gamma_\text{tot} = \sum_{i,j=1}^{N}\Gamma_{ij} = \sum_{i,j=1}^{N}& \frac{2 \pi \mu_0^2 \omega^2}{\hbar}[\B{d}_{\text{A}} \cdot \tens{G}(\B{r}_{\text{A}},\B{r}_{\text{D}i},\omega) \cdot \B{d}_{\text{D}j}]\notag \\\times &[\B{d}_{\text{D}i} \cdot \tens{G}^*(\B{r}_{\text{D}j},\B{r}_{\text{A}},\omega) \cdot \B{d}_{\text{A}}]
\end{align}
which reduces to Eq.~\eqref{GammaSingle} if $N=1$. Equation \eqref{Gammaij} gives the energy transfer rate for an excitation spread over $N$ donors arranged in an arbitrary way, and is the main expression we will be using for the rest of this work.

As a point of comparison, we will consider a convenient representation for a cloud of atoms which can be considered as identical. These are the Dicke states, based on the algebra of angular momentum. The general Dicke state for an $N$-particle system with $M$ excitations can be found from the completely excited state $\ket{e}_i \otimes \ket{e}_2\cdots\ket{e}_N \equiv \ket{E}$ via
\begin{equation}\label{DickeState}
 \ket{J;M} = \sqrt{\frac{(J+M)!}{N!(J-M)!}} \hat{J}_-^{J-M} \ket{E}
\end{equation}
where $J = \frac{N}{2}$, $M$ runs from $-J$ to $J$ in integer steps and  $\hat{J}_-$ is the collective angular momentum lowering operator $
\hat{J}_- = \sum_{i=1}^N \hat{\sigma}^-_i = \sum_{i=1}^N \ket{g}_i \bra{e}_i
$
If the system has one excitation, we have $M=-J+1$. Using this in \eqref{DickeState}, we get;
 $
 \ket{J;-J+1} = \sqrt{1/N!(N-1)!} \hat{J}_-^{N-1} \ket{E},
$
from which it follows that the matrix element of the collective dipole operator
$
\hat{\B{D}} = \sum_i \hat{\B{d}}_i = \B{d}_i(\hat{J}_- + \hat{J}_+)
$
for decay to the ground state $M=J$ is given by;
$
\bra{J,-J} \hat{\B{D}} \ket{J,-J+1} = \B{d}_i \sqrt{N}
$
Taking  $\B{d}_{\text{D}}\to \B{D} $ in \eqref{GammaSingle}, we then have
\begin{equation}\label{GammaSR}
\Gamma_\text{SR} = N^2 \Gamma
\end{equation}
which is also what we find when using Eq.~\eqref{Gammaij} with $\Gamma_{ij} = \Gamma$. If we instead carry out an incoherent sum ($\Gamma_{ij} = 0$ for $i \neq j$) over the $N$ donor-acceptor pairs, we get
\begin{align}\label{Gamma0}
\Gamma_{0} = \sum_{i=1}^{N}\Gamma_{i} = N \Gamma
\end{align}
This leads to a natural definition of the superradiant fidelity $F$:
\begin{equation}
F = \frac{\Gamma_\text{SR}}{N \Gamma_0} =  \frac{ \sum_{i,j=1}^{N} \Gamma_{ij} }{ N \sum_{i=1}^{N} \Gamma_{ii}}
\end{equation}
For the case of ideal (Dicke) superradiance have $F=1$, and for no superradiance (no coherence) we have $F = 1/N$. 

\begin{figure}
\includegraphics[width=\columnwidth]{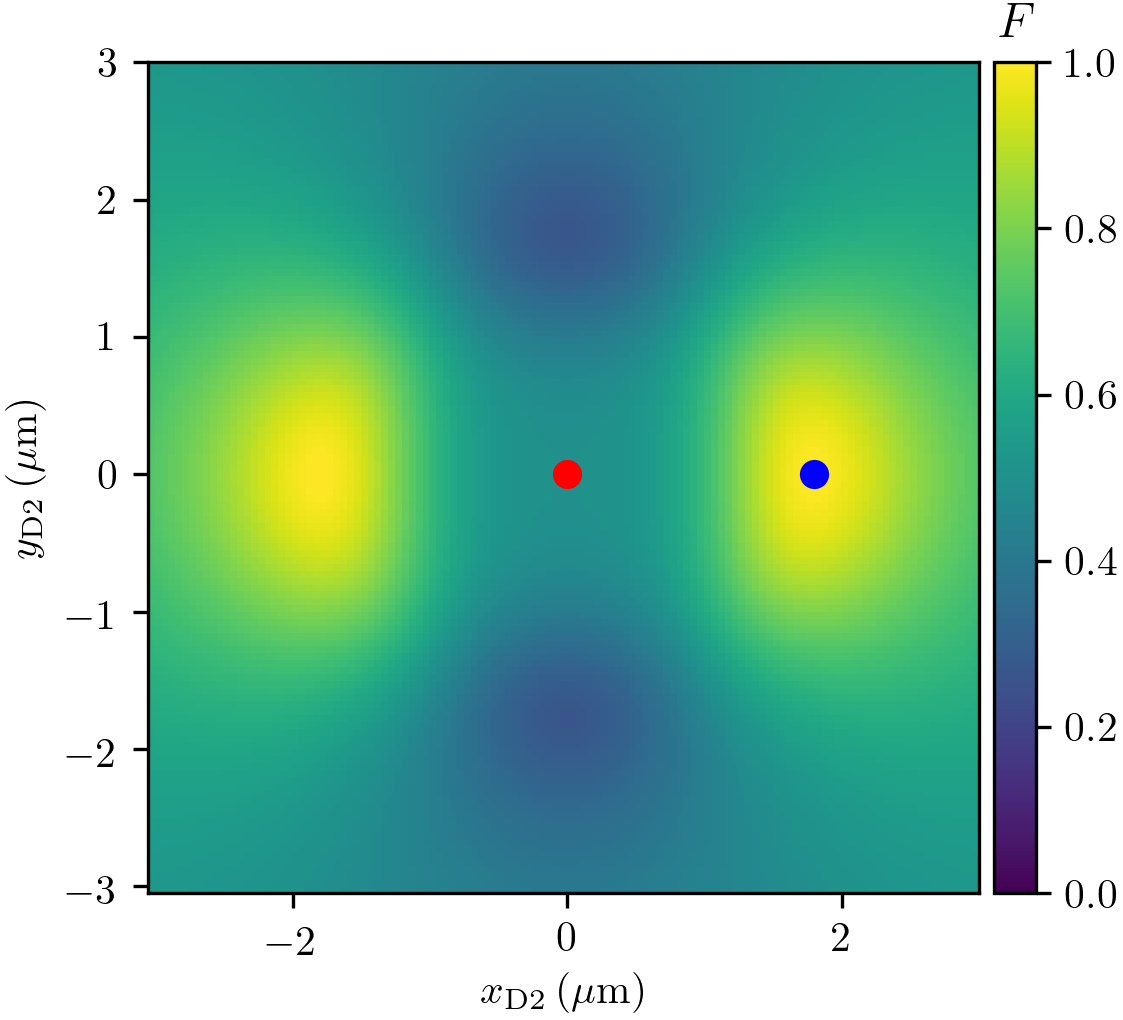} 
\caption{Superradiant fidelity $F$ for a two-donor system. The acceptor (red)  is placed at the origin, while the first donor (blue) is fixed $1.8\mu$m away. The second donor is free to move, with the colour bar representing the superradiant fidelity for a donor placed at that point. Both donors have transition wavelengths of $19\mu $m, meaning this example is in the non-retarded (electrostatic) regime.}\label{DiscreteDonors2}
\end{figure}

We can now calculate the fidelity in a variety of different situations. The simplest and most instructive is to consider a setup with two donors. We will assume that both donors and the acceptor are randomly oriented, under which conditions we can take $\B{d} \otimes \B{d} \to {|\B{d}|^2\mathbb{I}}/{3}$, meaning Eq.~\eqref{Gammaij} simplifies to;
\begin{align}\label{GammaIso}
\Gamma_{ij} &= \frac{2 \pi \mu_0^2 \omega^2}{9\hbar}|\B{d}_\text{A}|^2|\B{d}_\text{D}|^2\notag \\
& \quad \times \text{Tr} \left[ \tens{G}(\B{r}_{\text{A}},\B{r}_{\text{D}i},\omega)\cdot \tens{G}^*(\B{r}_{\text{D}j},\B{r}_{\text{A}},\omega) \right]
\end{align}
We will use the vacuum Green's tensor, given by
\begin{equation}\label{GVac}
\tens{G}(\B{r},\B{r}',\omega) = -\frac{c^2 e^{i\omega \rho/c}}{4\pi \omega^2 \rho^3}\left[f\left(\frac{\omega \rho}{c}\right)\mathbb{I}-g\left(\frac{\omega \rho}{c}\right) \B{e}_\rho\otimes \B{e}_\rho\right]
\end{equation}
where $\rho = |\bm{\rho}| = | \B{r}-\B{r}'|$, $\B{e}_\rho = \bm{\rho}/\rho$,  $f(x) = 1-\i x -x^2$, $g(x) = 3-3\i x -x^2$ and we have dropped a term proportional to $\delta(\B{r}-\B{r}')$ since the donor and acceptor points are never coinciding. 
Using this in Eq.~\eqref{GammaIso}, we find the results shown Fig.~\ref{DiscreteDonors2}, 
where it is clear that optimal placement of the second donor relative to the first is not limited to simply being near it as is usually assumed in Dicke physics, rather there is another region on precisely the other side of the acceptor for which the fidelity is maximised. Placing the second donor there, then subsequently placing a third donor at the new position of maximum fidelity and so on, one obtains that the optimal arrangement for SR-RET is \emph{two} clouds, one each side of the donor, which we shall investigate in more detail later.

\begin{figure}
\includegraphics[width=\columnwidth]{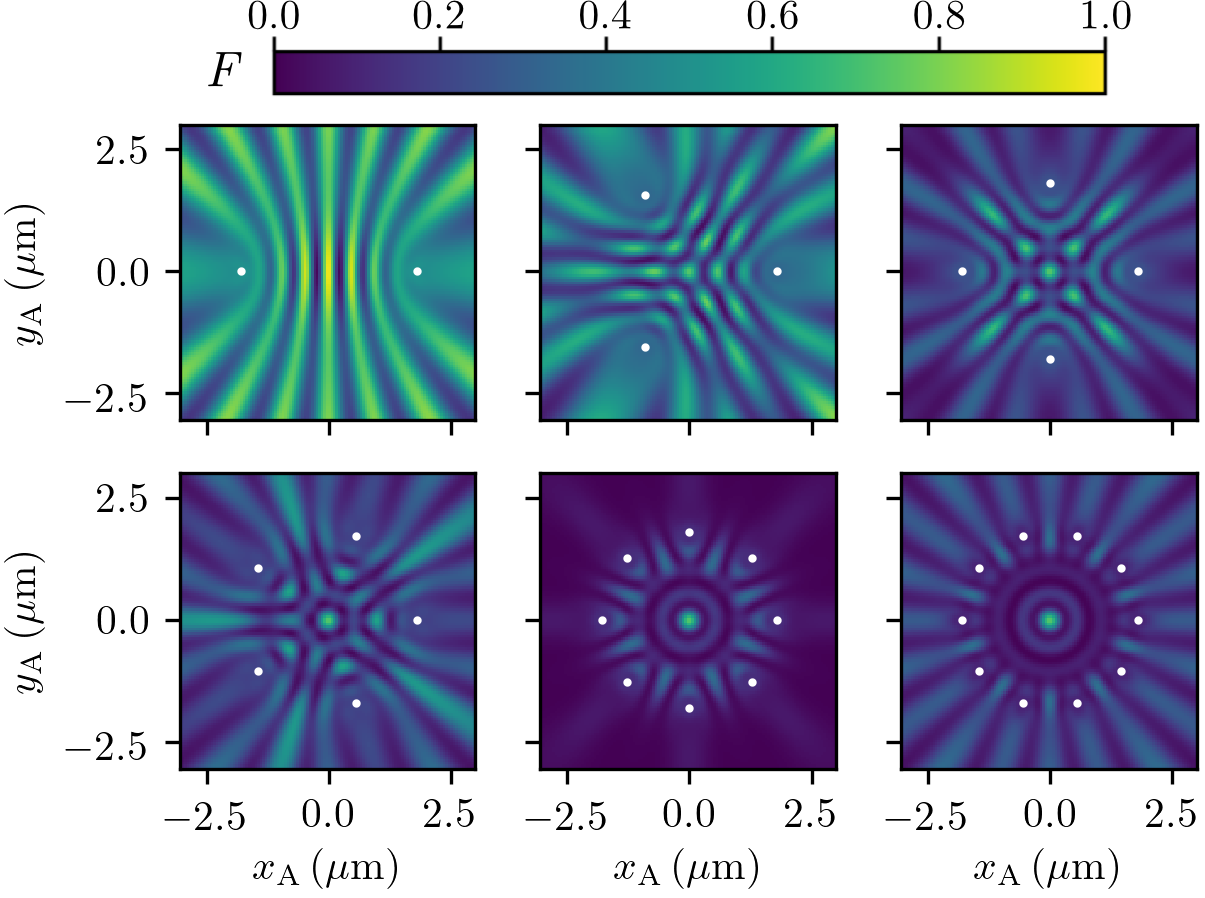} 
\caption{Superradiant fidelity $F$ for various numbers of donors ($N=2,3,4,5,8$ and $10$) arranged in a circle with a movable acceptor. The transition wavelength is 942nm, meaning this example is in the retarded regime, showing its characteristic oscillatory behaviour. }\label{DiscreteDonorsRET}
\end{figure}

Interestingly, grouping donors into three or more regions at equal distances from the donor does \emph{not} increase the fidelity. To see this, we evaluate the fidelity for donors evenly placed around a circle of radius $R$, with the acceptor at the centre. We find;
\begin{align}
F =& \frac{1}{4 N^2 \left(X^4+X^2+3\right)}  \Big[ N^2(3+X^2+3 X^4)\notag \\
&+\left(9+3 X^2+X^4\right)\sum_{\theta_i,\theta_j} \cos (2 \theta_i-2\theta_j)\Big]
\end{align}
where the angular sum runs over the positions of the $n$ donors, and $X = R \omega/c$. This expression demonstrates the distinctiveness of the two-donor case $N=2$. In that case $\theta_i$ and $\theta_j$ can both be either $0$ or $\pi$, meaning the cosine factor always evaluates to unity, so that the sum evaluates to $4$. The entire expression then simplifies to $F=1$. However, for all $N\geq 3$ the sum evaluates to zero, giving;
\begin{equation}
F =\frac{3 X^4+X^2+3}{4 \left(X^4+X^2+3\right)}
\end{equation}
which varies from $1/4$ as $X\to0$ (non-retarded regime) to $3/4$ as $X\to\infty$ (retarded regime). This is demonstrated in Fig.~\ref{DiscreteDonorsRET},
where we plot the fidelity for various numbers of donors arranged in a circle, with parameters chosen such that $X=12$. Increasing the number of donors, the fidelity for an acceptor placed at the center drops from its value $1$ for two donors to a constant value $0.746$ for any number of donors greater than two. This is close to the optimal value $3/4$ reached for larger $X$. Most crucially, adding more donors does not change the fidelity for a donor at the center, but merely reduced the number of locations where this maximal fidelity can be found in a focussing-like effect. 

The fidelity for an arbitrary position-dependent donor number density $n(\B{r})$ is;
\begin{align}
\Gamma_\text{SR}= \frac{2 \pi \mu_0^2 \omega^2}{\hbar}& \int \d^3 \B{r}_{\text{D}i}  \int \d^3  \B{r}_{\text{D}j} \,\,n(\B{r}_{\text{D}i})n(\B{r}_{\text{D}j}) &\notag \\\times &[\B{d}_{\text{A}} \cdot \tens{G}(\B{r}_{\text{A}},\B{r}_{\text{D}i},\omega) \cdot \B{d}_{\text{D}j}]\notag \\
&\times[\B{d}_{\text{D}i} \cdot \tens{G}^*(\B{r}_{\text{D}j},\B{r}_{\text{A}},\omega) \cdot \B{d}_{\text{A}}]
\end{align}
Simple expressions can be obtained for two homogenous clouds of donors of radius $R$ in the non-retarded (electrostatic) regime, obtained by letting $f(x)\to 1 $ and $g(x) \to 3$ in Eq.~\eqref{GVac}. For an acceptor placed at the origin with identical spheres of donors placed at $z = \pm z_0$ (with $z_0>R$), one finds;
\begin{equation}
F  = \frac{1}{{z_0^6}}(z_0-R)^3(z_0+R)^3
\end{equation}
The same expression is found for the case of a single sphere placed at $z=z_0$. Comparing the case of two spheres of radius $R$ to a single sphere with radius $2^{1/3}R$ (i.e. that with twice the volume), one finds that the fidelity is higher in the two-sphere case, as suggested by the case of discrete donors. 
\begin{figure}
\centering
\includegraphics[width=\columnwidth]{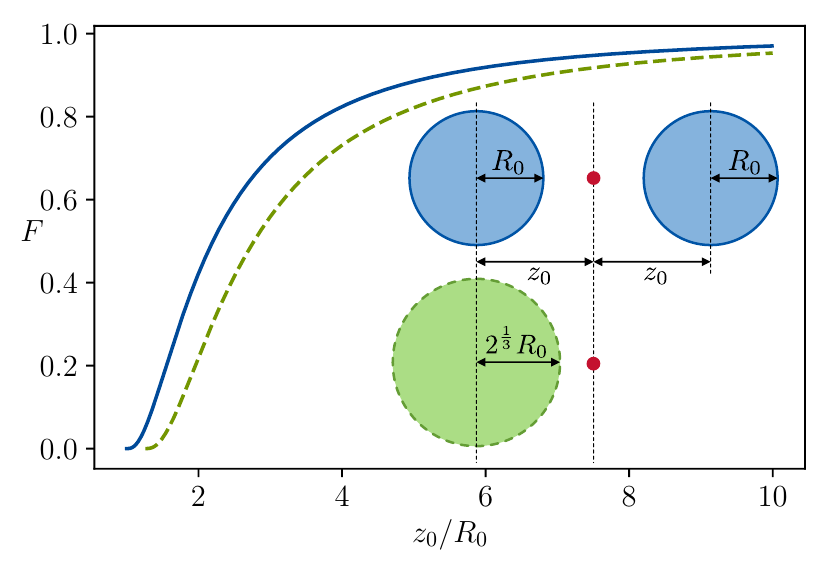} 
\caption{Superradiant fidelity as a function of separation for two spheres of radius $R_0$ placed either side of the acceptor (solid line), compared to a single sphere of the same total volume.}\label{SpherePlots}
\end{figure}

\begin{figure}[h!]
  \centering
\includegraphics[width=\columnwidth]{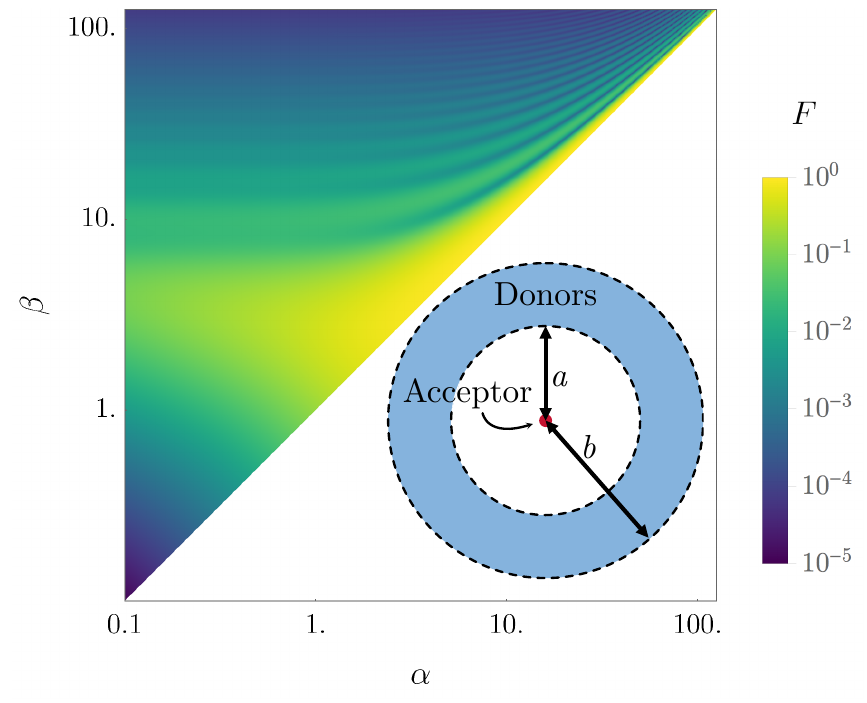}
\caption{Superradiant fidelity $F$ for an acceptor placed at the centre of a spherical shell of inner radius $a=\alpha c/\omega$ and outer radius $b = \beta c/\omega$}
\label{AlphaBetaPlot}
\end{figure}

The final example we consider is a hollow spherical cloud of donors with the acceptor at its center. Defining the inner and outer radius as $a$ and $b$ respectively and using the full Green's tensor \eqref{GVac}, we find;
\begin{align}\label{FSphere}
F &= \frac{16 \alpha  \beta  }{\pi ^2 (\alpha -\beta )^2 \left(\alpha  \beta  \left(\alpha ^2+\alpha  \beta +\beta ^2+3\right)+9\right)}\notag \\
&\qquad \times \Big(\alpha ^2 +\beta ^2+2+2 (\beta -\alpha ) \sin (\alpha -\beta )\notag \\
&\qquad \qquad -2 (\alpha  \beta +1) \cos (\alpha -\beta )\Big)
\end{align}
where $\alpha = a \omega/c$ and $\beta = b\omega/c$ ($\alpha < \beta$). As shown in Fig.~\ref{AlphaBetaPlot},
in the non-retarded regime (lower left corner), the rate (and consequently the fidelity) \emph{vanishes}, as can be seen by taking the limit $\alpha \to 0$ in Eq.~\eqref{FSphere}. This is a conclusion which holds even if the acceptor is placed off-center within the spherical shell. That this must be true can be seen by noticing that $\tens{G}(\B{r}_{\text{A}},\B{r}_{\text{D}i},\omega)$ is, in the non-retarded limit, the electrostatic Green's function. Integrating $\B{r}_{\text{D}i}$ over a sphere enclosing the observation point $\B{r}_{\text{A}}$ must then give zero due to the shell theorem of electrostatics \cite{Jackson1975}. Even beyond the retarded regime the  rate vanishes in the limit $\beta\to\infty$, corresponding to a spherical void inside an infinite bulk material. This demonstrates that resonance energy transfer will be strongly suppressed for an acceptor placed symmetrically deep inside a bulk medium of coherently excited donors.
The largest fidelity that can be found is in the retarded regime (upper right corner) when $\alpha \approx \beta$, corresponding to a thin shell. In this limit one finds from Eq.~\eqref{FSphere} that $F = {16}/(3\pi^2)\approx 0.54$.  

In this work we have shown that superradiant resonant energy transfer behaves qualitatively differently from superradiant spontaneous decay, relaxing the condition that the donors should be close to one another. Instead, for a collection of donors coherently sharing an excitation, superradiant energy transfer only requires these to occupy equivalent positions with respect to the donor. In this case, superradiant energy transfer takes place although the spontaneous decay in the absence of the acceptor is not superradiant. Quantifying the degree of superradiant energy transfer by means of a position-dependent fidelity, we have investigated the spatial conditions for superradiance in more detail. We have shown that the optimal geometrical arrangement is a dumbbell structure with donors placed in two clouds at opposite sides of the acceptor. On the contrary, circular donor placement quickly saturates the enhancement for more than two donors while spherically symmetric arrangements can even lead to a complete suppression of resonance energy transfer.

While our results remain valid for multiple acceptors in place of multiple donors, the simultaneous case of multiple donors and acceptors offers a richer parameter space that remains to be explored. In addition, we will in future work consider the robustness of our results towards imperfect coherence and the role the intial collective donor state in facilitating superradiant resonance energy transfer.

\end{document}